\documentclass{elsart1p}

\usepackage{graphicx}

\usepackage{amssymb}

\begin{document}

\begin{frontmatter}

\title{Interpretation of recent JLab results on quasi elastic (e,e'p) reactions off few-nucleon
systems}

 \author[label1]{C. Ciofi degli Atti}
 \address[label1]{Department of Physics, University of Perugia and Istituto Nazionale di Fisica
 Nucleare, Sezione di Perugia, Via A. Pascoli, I-06123, Italy }
 \ead{ciofi@pg.infn.it}
 \author[label1,label2]{L. P. Kaptari}
 \address[label2]{Bogoliubov Lab. of Theoretical Physics, 141980, JINR, Dubna, Russia}
 \ead{kaptari@thsun1.jinr.ru}
 \author{H. Morita}
 \address{Sapporo Gakuin University, 11 Bunkyo-dai, Ebetsu, Hokkaido 069-8555, Japan}
 \ead{hiko@sgu.ac.jp}

\begin{abstract}
  Recent JLab experimental data on quasi elastic $^3He(e,e'p)^2H(pn)$ and $^4He(e,e'p)^3H$ processes
   are interpreted using an approach based upon realistic wave functions and Glauber multiple scattering
   theory within a generalized eikonal
  approximation (GEA). The results of a non factorized  calculation of the left-right asymmetry
  $A_{TL}$ of the process $^3He(e,e'p)^2H$, obtained  using the full covariant form of the electromagnetic
  operator, are  also presented.

\end{abstract}

\end{frontmatter}

\section{Introduction}
\label{Intro}
 The exclusive process A(e,e'p)B in few-nucleon systems is one of the most promising
 tools to investigate  proton propagation in the medium. We have analyzed this process on
 the basis of realistic wave functions and a generalized eikonal approximation \cite{GEA1,GEA2}
  to describe the final state interaction. In this contribution we will present the results
  of our interpretation
 of  recent JLab data on the quasi elastic reactions   $^3He(e,e'p)^2H(pn)$ and $^4He(e,e'p)^3H$.

\section{Formulation}
 \label{Formulation}
 Within the factorization approximation, the differential cross section for the quasi-elastic
 A(e,e'p)B reaction
 can be written as follows
\begin{equation}\label{sigma}
    \frac{d^6\sigma}{d\omega d\Omega_edp
    d\Omega_p}=K\sigma_{ep}S_D(\mathbf{p}_m,E_m),
\end{equation}
where $K$ is a kinematical factor, $\sigma_{ep}$ the
electron-proton cross section and $S_D(\mathbf{p}_m,E_m)$  the
nuclear spectral function distorted by the final state interaction
(FSI), which is treated here within the
 generalized eikonal approximation (GEA)\cite{GEA1,GEA2}. In the
 case of few-nucleon system $S_D(\mathbf{p}_m,E_m)$ can be
 calculated using realistic wave functions and experimentally determined nucleon-nucleon scattering
 amplitude \cite{GEA2,He3,He4}. The violation of the factorization
 approximation will be discussed in the next Section.

\section{Results}
 \label{Results}
\begin{figure}[htb]
\begin{minipage}[htb]{65mm}
\hskip -0.18cm
  \includegraphics[width=6.5cm]{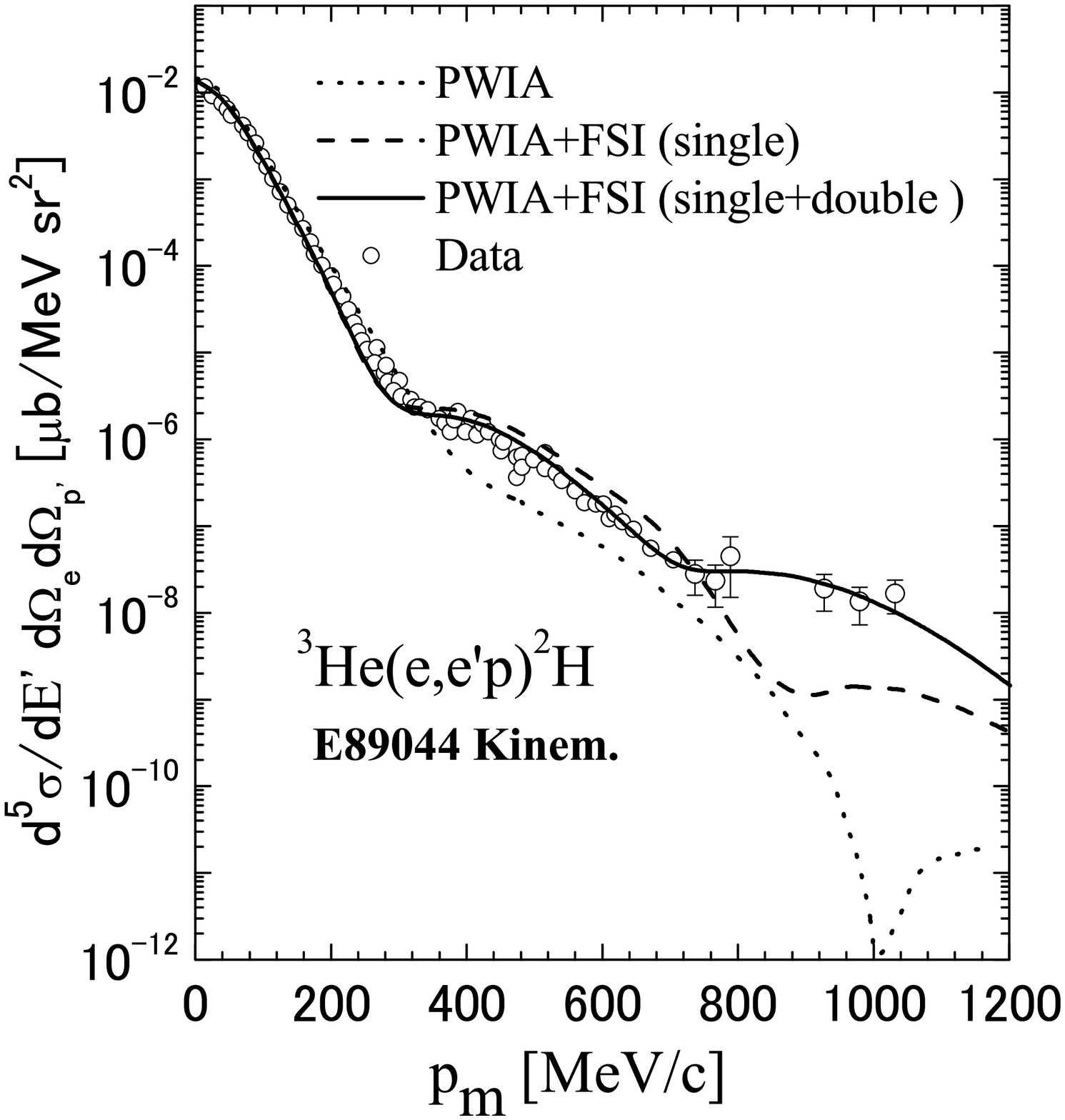}\\
  \vskip -1.0 cm
  \caption{Results for the $^3He(e,e'p)^2H$ reaction {\it vs} the missing momentum $p_m$ \cite{He3}.
Dots: PWIA ; dash:  FSI (single rescattering); full: FSI (single
plus double rescattering). Experimental data from \cite{He3data}.}
\label{fig:3He2bb}
\end{minipage}
\hskip 0.5cm
\begin{minipage}[htb]{65mm}
\hskip -0.25cm
 \includegraphics[width=5.8cm]{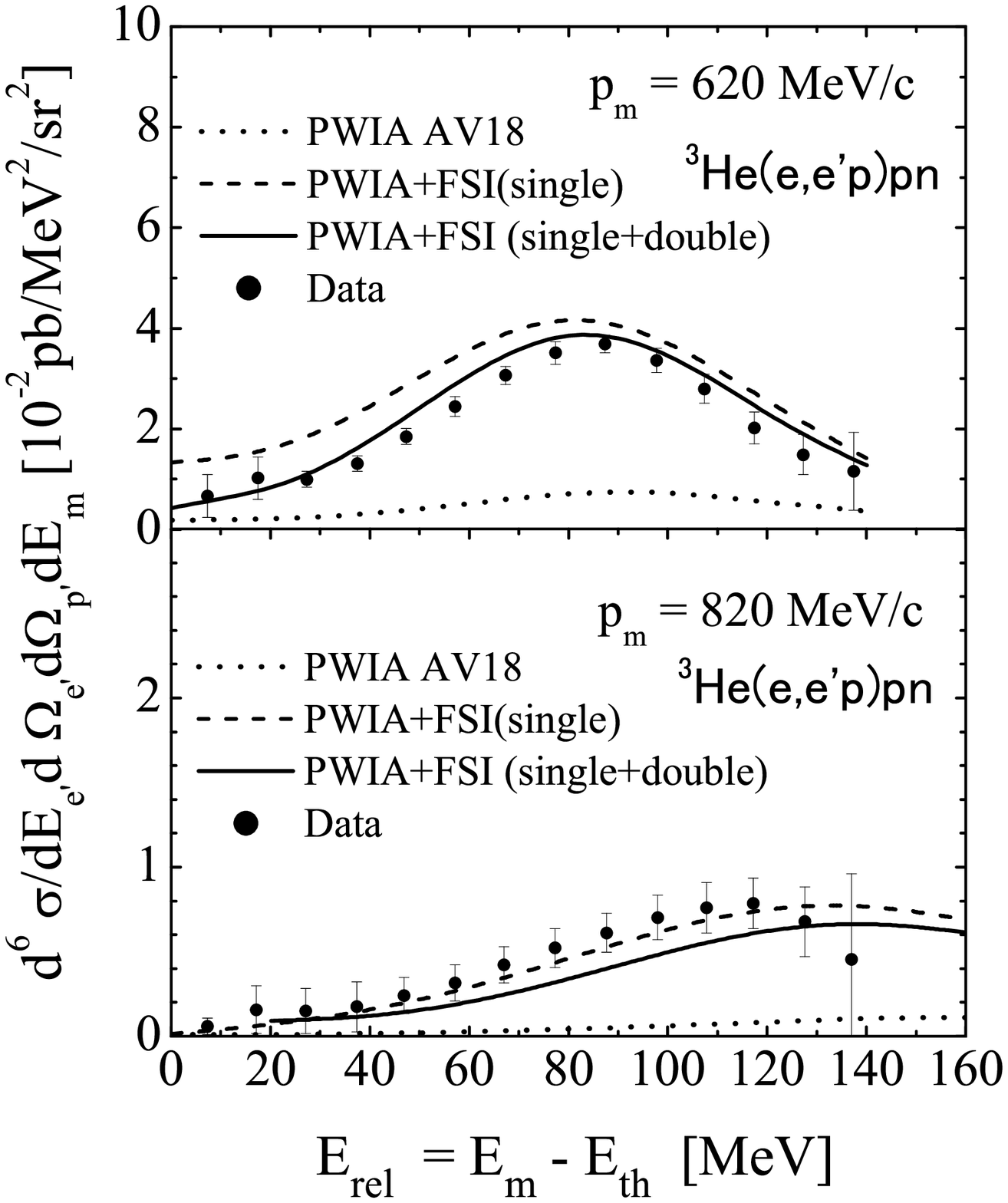}\\
\vskip -0.9cm
 \caption{The same as in Fig. \ref{fig:3He2bb} but for
the process $^3He(e,e'p)pn$ {\it vs} the excitation energy
$E_{rel}$ of the $n-p$ pair ($E_{th} =E_3$ is the two-nucleon
emission threshold in $^3He$).} \label{fig:3He3bb}
\end{minipage}
\end{figure}

The results for the $^3He(e,e'p)^2H(pn)$ reactions \cite{GEA2,He3}
are presented in Fig.\ref{fig:3He2bb} and Fig.\ref{fig:3He3bb}
respectively; a very good agreement between the GEA calculation
and the experimental data can be seen.
 Fig. \ref{fig:3He2bb} shows that the cross section exhibits different slopes, which corresponds
 to the PWIA and to the single- and double-rescattering contributions,  respectively
  \cite{He3}.

\begin{figure}[htb]
\begin{minipage}[htb]{65mm}
\vskip -0.7cm \hskip -0.18cm
  \includegraphics[width=6.5cm]{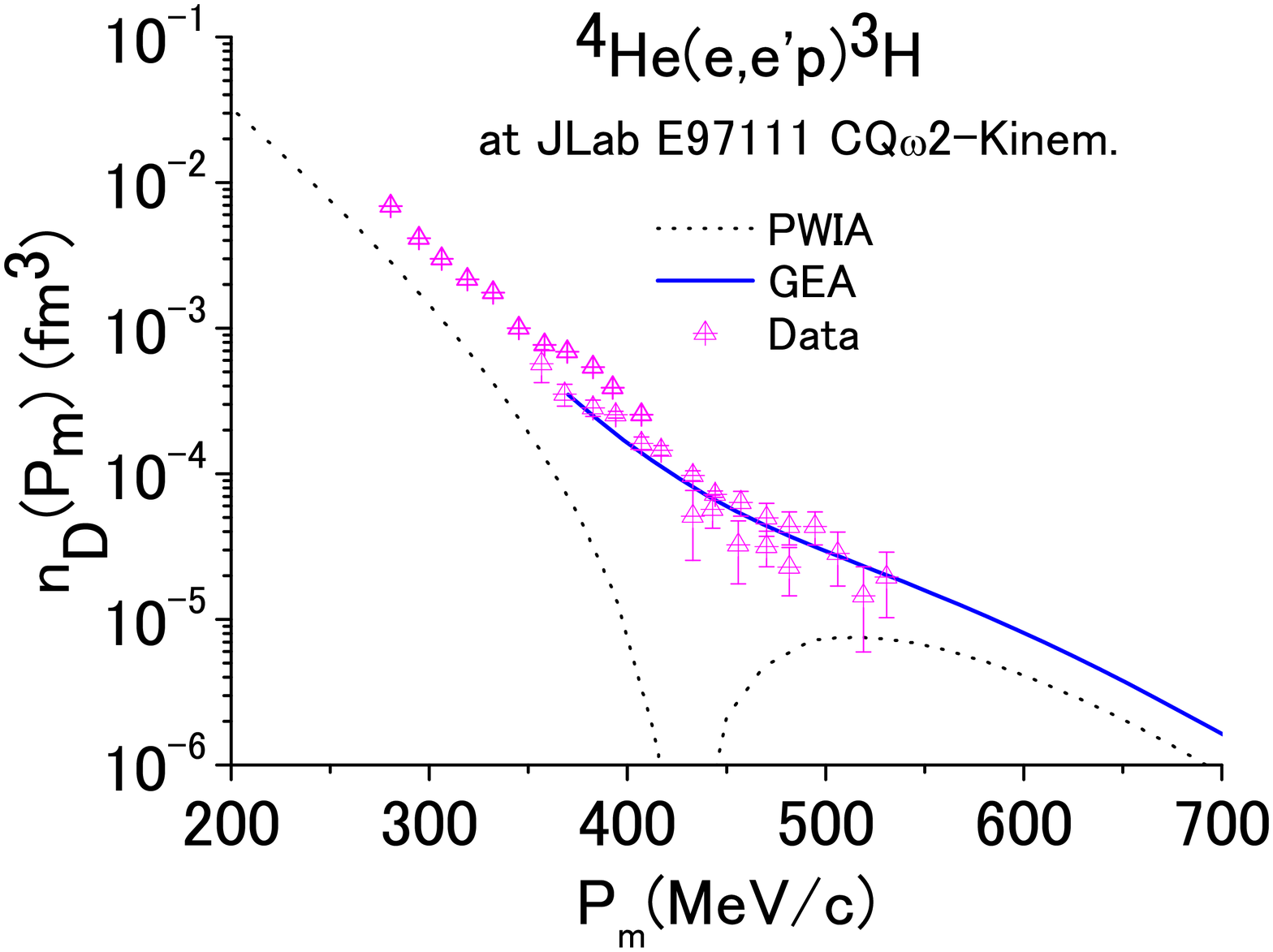}\\
  \vskip -0.7cm
  \caption{The reduced cross section
 $n_D(\mathbf{p}_m)=[d^5\sigma/(d\nu d\Omega_{e}d \Omega_p)]\times
[{\mathcal K}\sigma_{ep}]^{-1}$  for the process  $^4He(e,e'p)^3H$
\cite{He4}. Preliminary experimental data from \cite{He4data}.}
\label{fig:4Hecqw2}
\end{minipage}
\hskip 0.5cm
\begin{minipage}[htb]{65mm}
\vskip -0.1cm \hskip -0.5cm
 \includegraphics[width=6.5cm]{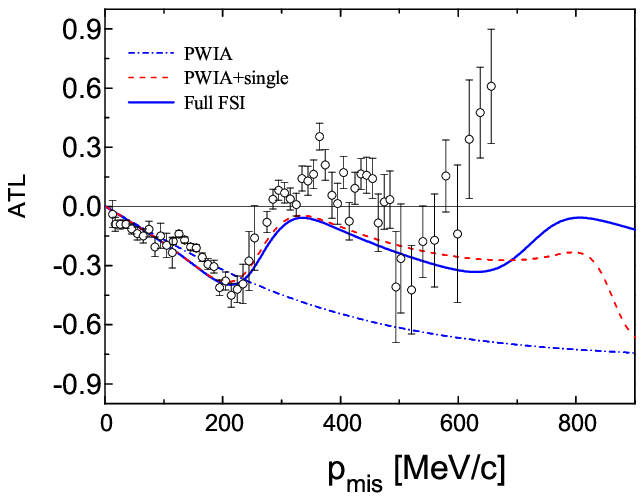}\\
  \vskip -0.9cm
\caption{The  asymmetry $A_{TL}$ for the process $^3He(e,e'p)^2H$
\cite{nonfact}. Dot-dash:  PWIA;  dashed and full: nonfactorized
results corresponding to single- and double-rescattering FSI,
respectively. Experimental data from \cite{He3data}}.
\label{fig:ATL}
\end{minipage}
\end{figure}

Our  results \cite{He4} for the $^4He(e,e'p)^3H$ process are shown
in Fig. \ref{fig:4Hecqw2}, and again a satisfactory agreement with
the experimental data can be seen.

Finally in Fig. \ref{fig:ATL}  the left-right asymmetry $A_{TL}$

\begin{equation}\label{ATL}
    A_{TL}=\frac{\sigma(\phi=0^{\circ)}-\sigma(\phi=180^{\circ)}}{\sigma(\phi=0^{\circ)}+\sigma(\phi=180^{\circ)}},
\end{equation}
calculated within a non-factorized  momentum space approach
,without using any kind of non-relativistic reduction, is shown
Fig. \ref{fig:ATL} \cite{nonfact}; it can be seen that
 the factorized calculation cannot describe the data at  $p_m\geq 200 (MeV/c)$,
 whereas the
non-factorized calculation can reproduce the data up to $p_m
\simeq 600 (MeV/c)$ , but fails at higher momenta. The extension
of the non factorized approach to the process $^4He(e,e'p)^3H$  is
in progress \cite{nonfact2}.

\section{Summary}
 \label{Summary}

We have analyzed recent JLab experimental data on quasi-elastic
processes off few-nucleon systems using realistic wave functions
and the Generalized Eikonal Approximation to treat  the  FSI. The
factorized calculations reproduces very well the three-body break
up and the right wing of of the two-body break. Our non-factorized
calculation of the asymmetry $A_{TL}$  does very well up to
$p_m\simeq600 (MeV/c)$, but a strong disagreement between theory
and experiment at higher momenta, common to many calculations,
remains to be explained.

\end{document}